\documentclass[journal,onecolumn]{article}
\usepackage{graphicx}          
\usepackage{epstopdf}           
\usepackage{float} 

\usepackage{circuitikz}
\usepackage[shortlabels]{enumitem}
\usepackage{tabularx, multirow}	
\usepackage{subcaption}
\usepackage[a4paper, total={7in, 9in}]{geometry}

\usepackage{amssymb}
\usepackage{amsmath}
\usepackage{amsfonts}
\usepackage{mathtools}
\usepackage{accents}

\usepackage{url}
\usepackage{caption,booktabs}
\newtheorem{thm}{Theorem}
\newtheorem{rem}{Remark}
\newtheorem{lem}{Lemma}
\newtheorem{pf}{proof}

\title{\LARGE \bf
	Integrated Supervised Adaptive Control for the More Electric Aircraft}

\author{Alberto~Cavallo,
	Giacomo~Canciello,
	Antonio~Russo
	\thanks{A. Cavallo, G. Canciello and A. Russo are with the Department
		of Engineering, Universit\`a degli Studi della Campania ``L. Vanvitelli'', Aversa(CE),
		Italy, 81032 e-mail: \{alberto.cavallo,giacomo.canciello,antonio.russo1\}@unicampania.it.}
	}

\begin{document}

%
%

%

\maketitle
\begin{abstract}                
The innovative concept of Electric Aircraft is a challenging topic involving different control objectives. For instance, it becomes possible to reduce the size and the weight of the generator by using the battery as an auxiliary generator in some operation phases. However, control strategies with different objectives can be conflicting and they can produce undesirable effects, even instability. For this reason an integrated design approach is needed, where stability can be guaranteed in any configuration. In other words,  the design of the supervisory controller must be interlaced with that of low-level controllers. Moreover, uncertainties and noisy signals require robust control techniques and the use of adaptiveness in the control algorithm. 
In this paper,  an aeronautic application aiming at recharging batteries and to use the battery to withstand generator overloads is addressed. Detailed and rigorous stability proofs are given for any control configuration, including the switching phases among different control objectives. Effectiveness of the proposed strategies is shown by using a detailed simulator including switching electronic components. 
\end{abstract}




\section{Introduction}
In July 2018 the flight of the Norway's transport minister in an aircraft completely powered by electric energy was the first step towards the decision of providing Norway with a fleet of electric planes by 2040\cite{dowling2018}.  A small airline based in Vancouver is currently working for commercial flights done by a purely electric aircraft to be operative in 2022~\cite{patel2019}. 

The above are just two examples of what is now considered the future of flight, i.e., the ``electric plane''~\cite{fut_flig}, an idea  that has been addressed since the beginning of the century, with  the preliminary concept of More-Electric Aircraft (MEA~\cite{mea2002}). A detailed review of opportunities and potential for electric aircraft, with a chronological presentation of the associated technologies, is presented in~\cite{GOHARDANI2011369}, with emphasis on the reduction of the empty weight of the aircraft, on the benefits in terms of safety and reliability (due to increased fault-tolerance), and on reduction of maintenance costs (due to reduction of moving parts).  Under the generic umbrella of ``Electric Aircraft''  different targets and opportunities are comprised. Major advancements are required in the field of electric battery enhancement, e.g., by using fuel cells~\cite{SEHRA2004199} to replace current gas turbine APU (Auxiliary Power Units) or even for fuelling electrically powered fan (E-Fan)~\cite{airbus_efan,tweed2014}. Other crucial points are  the increase of generator power density (in terms of power per unit of weight),  and in general, all the issues related to dramatic changes in  electric power generation, distribution and management~\cite{TowardsMEA}. 

The role, the opportunities and the needed development for electric machines and motors have been widely analysed~\cite{Weimer2003}, and the potential for improvement in this field is clearly acknowledged, see~\cite{Cao2012}, \cite{Sarlioglu2015} for  exhaustive discussions on this topic. In contrast to the well-assessed interest in electric system and power electronic field, the new requirements in the field of Automatic Control is not clearly pointed out. The standard approach to controller design for MEA applications  starts from a  two-layer based modelling phase, resulting in {\em functional} models and {\em behavioural} models~\cite{Bozhko2014}. 

Referring to the case of DC-DC power converters, that is of interest in this paper, the functional modelling produces  nonswitching averaged models, that are employed to assess system dynamics and stability, while behavioural models  are intended to model high-frequency phenomena, including commutation for switching devices. The next step in most cases, is to employ simple standard controllers (PI and PID) for low-level control~\cite{Chen2012}, while high-level strategies aiming at the interaction of different controlled devices are based on ingenuous heuristic approaches. This na\"\i f approach is reasonable  in the classic approach of the Power Electronic field, where the focus is on the topology of the converter and the modulation techniques~(see~\cite{HOSSAIN2018} and references therein for a broad analysis), while the details of the control law receive marginal attention since the PI usually suffices. In some cases, due to the simplicity of the functional model, only very rough simulations are carried out, and the  focus is moved to the experimental implementation (see, for instance~\cite{Zhang2010}, where different controllers are tested with this approach). This methodology is effective when single control tasks are considered, with only one working condition around which the controlled system evolves during time. But in the case of multiple tasks, this kind of approach is limited from the control point of view, because it is implicitly based on the linearised model of the system and is neither able to  deal with large variations of the variable, nor guarantee global or semiglobal stability. A more rigorous approach uses sliding mode control (both standard~\cite{sira_ram_book}, \cite{utkin_book} and second-order~\cite{Shtessel_Zinober_Shkolnikov_2003}). One of the difficulties faced with this approach is to produce a control signal that can be implemented with a  PWM modulation~\cite{Lai_Tse_2006}, that is the standard technology used in converter's drives. 

In addition, in the case of multiple tasks, when the operative point changes also the control must change.  Different from the above approach, from the automatic control point of view there is the need for integration  between the design of the low-level layer, basically controlling the single devices (power converters, motors, generators) and high-level control, defining how different controlled device interface each other when control objectives change. Indeed, a typical and well-known problem arise when two controlled devices behave like constant-power load, due to the action of the controller, and this can even result into instability~\cite{Emadi1999}. Even more subtle is the case of a supervisory control action autonomously switching among different stable configurations, e.g., due to changing control objective. Also in this case, counterexamples have been discovered showing that switching among stable configurations can result into overall instability~\cite{Branicky1998}.
The above drawbacks can be explicitly dealt with by co-designing high-level and low-level control. Specifically, low-level controllers must fulfil standard requirements (e.g., closed-loop stability, robustness) but they must also produce  an estimate of the region of attraction. On the other side, high-level control must reformulate control objectives so as to adapt their action  to the current condition of the controlled system, with reference to the regions of attraction mentioned above. 

Recently, a multi-objective problem for the MEA involving the control of a bidirectional DC-DC converter has been addressed by the authors. Specifically, the aircraft power grid can be modelled as a simple two-busbars system~\cite{tooley2009aircraft}, one high voltage (HV) DC bus at $270$V and a low voltage (LV) DC bus at $28$V, with a DC-DC bidirectional converter in between. The HV bus is energised by a starter-generator followed by a rectifier. On the HV side all the ``heavy'' loads are present (e.g., anti-icing and de-icing). On the LV side a battery is located, for avionics and emergency conditions.  In normal operating conditions the battery is charged by the generator (so the battery is a further load for the generator). Since generator sizing is based on its capability to withstand large loads for more than $5$s, (the so-called $5$s-$5$ min capability~\cite{CavalloAutomatica17b}) the idea is to use the battery to help the generator in the case of overload, so as to reduce generator sizing (and weight). It is clear from the above discussion that the battery must be able to supply power to the HV size within $5$s from the request, so as to comply with the  $5$s-$5$ min constraint. In~\cite{cavallo:ijrnc17}  a solution employing high-gain control has been proposed, with an approach with properties similar to the classic Integral Sliding Control~\cite{integral_utkin}, i.e., the state starts directly from a well-defined sliding manifold and remains in a neighbourhood of it  thereafter. Moreover, when the state is on the manifold the closed-loop system behaves like a {\em linear system}, with apparent advantages in terms of simplicity and global stability. However, the crucial point in this design is that the parameters of the system have to be exactly known in order to compute a suitable control gain. A way to overcome this limitation is use data-mining algorithms with a learning phase~\cite{cavallo_energ18}. Another approach, considered in this paper, is to use an adaptive approach to directly estimate this gain. The use of adaptive approaches is not new in  power converter control design. In~\cite{Pahlevaninezhad2012} an adaptive design is proposed for a DC/DC converter for electric vehicles. However, this approach in based only on steady-state consideration, hence it cannot deal with stability issues. A rigorous approach, based on output regulation and an asymptotic computation of the stabilising feedback, is presented in~\cite{KARAGIANNIS2003}, where only the case of a boost converter is considered. Also in~\cite{NIZAMI2016} an adaptive approach is proposed for the control of a buck converter, training the parameters of a  neural network until a stabilising control action is achieved.  The adaptive approach has proved to be effective when the load has to be estimated, as in~\cite{canudas2010}, where the control of a boost converter is addressed. Recently, in~\cite{BENEUX2019} an approach based on stability analysis of switched systems has been proposed and applied to a Flyback converter. Finally, an adaptive approach producing a sliding mode controller for a boost converter has been proposed in~\cite{Fadil2006}. All the above approaches focus on e single objective, e.g., a current or a voltage regulation. The strategy proposed in this paper focusses on two different objectives 

The rest of the paper is organised as follows. Section~\ref{sec:model} presents the model of the bidirectional converter to control, along with some physical considerations. Section~\ref{sec:control} is the core of the paper and presents the mathematical results. It is split into three different subsections, showing the uniform stability of the controlled system, the stability of the adaptive control laws with different control objectives, and the supervisory control, respectively. All the strategies are presented with a single, integrated design approach taking into account an estimate of the Regions of Attraction of the control as a vital part of the design. Section~\ref{sect:simul} presents a possible implementation of the integrated strategy discussed in Section~\ref{sec:control}, along with detailed simulation to test the effectiveness of the proposed. Finally, some conclusions are presented in Section~\ref{sec:conclusion}.
\section{BBCU Model}\label{sec:model}
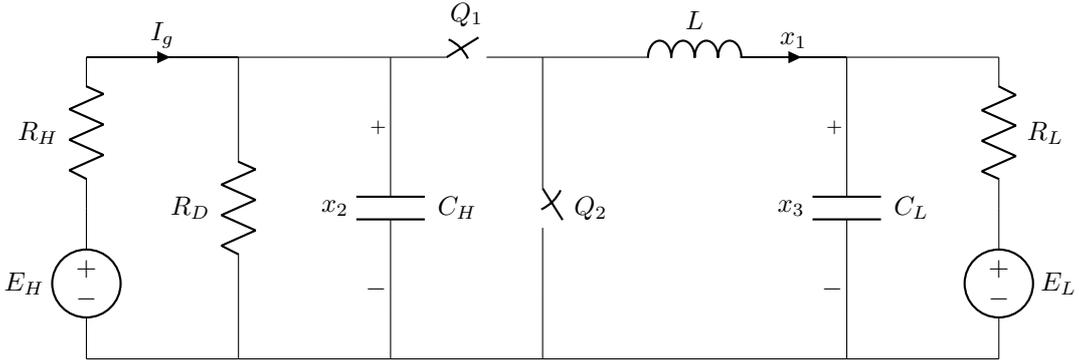
\begin{figure}[ht]
      \centering
      \ctikzset{bipoles/length=1.5cm}
      \begin{circuitikz} [scale=0.4, american]
    \draw (0,5) to[V, v_=$E_H$] (0,0);
    \draw (0,5) to[R=$R_H$] (0,10);
    \draw (0,10) to[short,i^=$I_g$] (5,10);
    \draw (5,0) to[R=$R_D$] (5,10);
    \draw (5,0) to[short](10,0);
    \draw (5,10) to[short](10,10);
    \draw (10,10) to[C=$C_H$,v_=$x_2$] (10,0);
    \draw (5,0) to[short](0,0);
    \draw (10,10) to[cspst=$Q_1$] (15,10);
    \draw (15,10) to[cspst=$Q_2$] (15,0);
    \draw (15,0) to[short](10,0);
    \draw (15,10) to[L=$L$,i^=$x_1$] (25,10);
    \draw (25,10) to[C=$C_L$,v_=$x_3$] (25,0);
    \draw (25,10) to[short] (30,10);
    \draw (30,10) to[R=$R_L$] (30,5);
    \draw (30,5) to [V, v=$E_L$] (30,0);
    \draw (30,0) to[short] (15,0);
     \end{circuitikz}
\caption{Bidirectional Buck-Boost converter schematic}
\label{fig:buck_boost_scheme}
\end{figure}

The BBCU bidirectional converted considered in this paper is shown in Figure~\ref{fig:buck_boost_scheme}. This circuit is representative of the HV and LV buses on-board aircraft. As stated in the Introduction, the HV side voltage source is a three-phases generator undergoing rectification which is here schematically represented as an ideal DC voltage generator $E_H$ and its internal resistance $R_H$. Also the battery on the LV side is represented by an ideal voltage generator $E_L$ and its internal resistance $R_L$. The bidirectional converter  employs an inductor $L$ and two capacitors, one for each side, $C_H$ and $C_L$ for the HV and the LV side, respectively. The circuit modulates  power due to   two switches, $Q_1$ and $Q_2$, which are controlled by an ON/OFF synchronised signal. Usually, with PWM implementations, the switching frequency and duty-cycle determine the power flow and its intensity. Since only active power is of interest for this application, the load can be modelled as a resistor $R_D$ with a slowly time-varying resistance value. Therefore, a load variation will be simulated by a variation of the circuit resistance $R_D$. 


The circuit equations are easily derived for both configurations ($Q_1$ OFF, $Q_2$ ON) and ($Q_1$ ON, $Q_2$ OFF) and can be written in a compact way as

\begin{eqnarray}
\dot{x}_1&=&-\frac{1}{L}x_3+\frac{1}{L}x_2 u\label{eq:buck_boost_model1} \\
\dot{x}_2&=&-\alpha x_2-\frac{1}{C_H}x_1 u+\beta_H\label{eq:buck_boost_model2} \\
\dot{x}_3&=&\!\frac{1}{C_L}x_1-\frac{1}{R_L C_L}x_3+\beta_L\label{eq:buck_boost_model3} 
\end{eqnarray}
where
\begin{equation}
\alpha= \frac{1}{R_{DH}C_H}, \quad R_{DH}=\frac{R_DR_H}{R_D+R_H},\quad \beta_i=\frac{E_i}{R_i C_i},\quad i\in\{H,L\},
\end{equation}
$x_1$ is the current flowing through the inductor $L$, $x_2$ is the voltage across the capacitor $C_H$ on the HV bus side, $x_3$ is the voltage across the capacitor $C_L$ on the LV bus side, and the control $u\in\{0,1\}$ is a binary variable defining  the two configurations. Note that $R_{DH}$  is physically the value of the resistance resulting from the parallel connection of $R_D$ and $R_H$.
Finally, we assume that $E_H>E_L$ and that the resistor $R_H$ is small enough so that 
\begin{equation}\label{eq:ELEH}
E_H>\left(1+R_H/R_D\right)E_L
\end{equation}
Physically this requirement is very reasonable, since usually $R_H/R_D\ll 1.$

\section{BBCU Control}\label{sec:control}

The control of the BBCU is designed in a hierarchical manner. A low-level control layer is designed in order to accomplish proper current tracking capabilities while a high-level control layer is responsible for the selection of the proper BBCU operating modality according to a prescribed functional policy. Preliminarily, we point out a structural property of the system under consideration. Then we will discuss the control strategies. 

\subsection{Uniform Stability of the BBCU}\label{sec:US}
In this Section we show that there is no loss of generality in assuming the BBCU to operate in a bounded region in the variables state space. To formally show this statement, we present the following Lemma.  

\begin{lem}[Uniform Stability]\label{lem:us}
Consider the system~(\ref{eq:buck_boost_model1})--(\ref{eq:buck_boost_model3}) and let $u$ be any  control $u\in{\mathcal{L}}_\infty$. Then the system~(\ref{eq:buck_boost_model1})--(\ref{eq:buck_boost_model3}) is uniformly stable.
\end{lem}
\begin{pf}
Consider a nonzero solution $\xi=\xi(t)$ of the  system~(\ref{eq:buck_boost_model1})--(\ref{eq:buck_boost_model3}) for a bounded $u=u(t)$. Using the change of variables 
\begin{equation}\label{eq:coord_change}
y(t) = x(t)-\xi(t)
\end{equation}
the translated system can be written as 
\begin{eqnarray}
\dot{y}_1&=&-\frac{1}{L}y_3+\frac{1}{L}y_2 u(t)\label{eq:tbuck_boost_model1} \\
\dot{y}_2&=&-\alpha y_2-\frac{1}{C_H}y_1 u(t)\label{eq:tbuck_boost_model2} \\
\dot{y}_3&=&\!\frac{1}{C_L}y_1-\frac{1}{R_L C_L}y_3\label{eq:tbuck_boost_model3} 
\end{eqnarray}
Note that, due to the assumption $u\in{\mathcal{L}}_\infty$, the right-hand side of~(\ref{eq:tbuck_boost_model1})--(\ref{eq:tbuck_boost_model3}) is locally Lipschitz in $y$ on a domain $D\subset\mathbb{R}^n$. Now, consider the Lyapunov function
\begin{equation}\label{eq:lyapy}
V(y) = \frac{1}{2}\left(Ly_1^2+C_Hy_2^2+C_Ly_3^2\right)
\end{equation}
whose time derivative along the trajectory of the system is
\begin{equation}\label{eq:dlyapy}
\dot{V}(y) = -\frac{1}{R_{DH}}y_2^2-\frac{1}{R_{L}}y_3^2\le 0.
\end{equation}
Thus, using~\cite[Theorem 4.8]{khalil} the uniform stability of $y=0$ is proved.
\end{pf}

Equipped with the above Lemma, it makes sense to assume that all the variables are bounded, and in particular there exist  $X_1^-$ and positive scalars $X_1^+$, $X_2^-$, $X_2^+$,  $X_3^+$, such that
\begin{eqnarray}
	&&X_1^- \leq x_1 \leq X_1^+, \label{eq:Th1_hp1}\\
	&&X_2^- \leq  x_2   \leq X_2^+, \label{eq:Th1_hp2}\\
	&&0 <  x_3   \leq X_3^+, \label{eq:Th1_hp3}
\end{eqnarray}
Strictly speaking, the hypothesis of positivity of $x_2$ and $x_3$ does not come directly from Lemma~\ref{lem:us}, but is a trivial requirement in this kind of applications, and is physically sound. 

Moreover, although the lemma considers the case of time-varying control $u$, in practical implementations  the control always approaches constant values, since stepwise constant references are used. Then it will be useful to denote by $x_u^*(\alpha) = \left[ x_{1u}^*(\alpha) \quad x_{2u}^*(\alpha) \quad x_{3u}^*(\alpha)\right]^T $ the steady-state solution of the system~(\ref{eq:buck_boost_model1})--(\ref{eq:buck_boost_model3}) with {\em any} fixed $u$ in the interval  $[0,1]$ and for a given $\alpha$ in~(\ref{eq:buck_boost_model2}).\footnote{For the sake of notational simplicity we will drop the dependence of the solution on $\alpha$ hereafter.}  

It is of particular interest the investigation of stability of the steady-state solutions  in both extreme configurations (namely $u\equiv0$ and $u\equiv1$, fixed). It is clear that in both configurations, system~(\ref{eq:buck_boost_model1})--(\ref{eq:buck_boost_model3}) reduces to an LTI system. Specifically, for $u\equiv 0$, the system has a globally exponentially stable 
equilibrium point at 
\begin{eqnarray}
x^*_{10} &=& -\frac{E_L}{R_L} \nonumber\\
x^*_{20} &=& \frac{R_{DH} }{R_H}E_H\label{eq:eqpoint_1}\\
x^*_{30} &=& 0. \nonumber
\end{eqnarray}

Analogously, when $u\equiv 1$, the dynamic matrix is 
\begin{equation} \label{eq:dyn_mat1}
\begin{bmatrix}
0 & \frac{1}{L} & -\frac{1}{L}  \\
-\frac{1}{C_H} & -\alpha & 0 \\
\frac{1}{C_L} & 0 & -\frac{1}{R_LC_L}   
\end{bmatrix},
\end{equation}
that is clearly Hurwitz (it is sufficient to apply the Routh-Hurwitz criterion) and the globally exponentially stable equilibrium point is at
\begin{eqnarray}
x^*_{11} &=& \frac{R_{DH}/R_H\,E_H-E_L}{R_{DH}+R_L}\nonumber\\
x^*_{21} &=& \frac{R_{DH}R_L}{R_{DH}+R_L}\left( \frac{E_H}{R_H} + \frac{E_L}{R_L} \right)   \label{eq:eqpoint_2}\\
x^*_{31} &=& x^*_{21}. \nonumber
\end{eqnarray}
These properties will be used in the next Section.

\subsection{Low-Level Control}\label{sec:lowlevel}
Both the battery current control and the generator current limitation are based on the definition of a sliding manifold 

\begin{equation}\label{eq:slid_man}
{\mathcal S}= \left\{(k,x)|\sigma(k,x)=0, \forall t\ge 0\right\} 
\end{equation}
where  the sliding function $\sigma(k,x)$ is
\begin{equation}\label{eq:slid_fun}
\sigma(k,x)=k(t)x_2-x_1 
\end{equation}
where $k$ is a design parameter. The basic idea is that when $\sigma(k,x)=0$, i.e., the state is on the manifold ${\mathcal S}$, the control objective has been achieved for a suitable value $k$ chosen adaptively. 
A similar approach to this problem has already been proposed in~\cite{cavallo:ijrnc17} through High-Gain  Control Theory and Tikhonov's Theorem on the infinite time horizon. We here just recall the final results of~\cite{cavallo:ijrnc17} since they will be used later in this paper. The key point is that the parameter $k(t)$ has to be chosen so that it asymptotically approaches the value $k_{\infty,1}$,  
\begin{equation}\label{eq:k_old1}
k_{\infty,1}=\frac{E_H-\sqrt{E_H^2-4R_H^2C_H\bar{x}_1(R_L\bar{x}_1+E_L)\alpha}}{2R_H(R_L\bar{x}_1+E_L)}
\end{equation}
in the case of battery recharging, assuming a constant current recharge, i.e., a constant reference $\bar{x}_1$ for the inductor current.  In the case of generator current limitation,   $k(t)$ must approach
\begin{equation}\label{eq:k_old2}
k_{\infty,2}=\frac{-E_L+\sqrt{E_L^2-4R_L\bar{x}_{2}(\alpha C_H\bar{x}_{2}-E_H/R_H)}}{2R_L\bar{x}_2}
\end{equation}
where $\bar{x}_2$ is a prescribed voltage reference such that the generator current does not go beyond an upper bound (see later). 
However, the control strategy proposed above presents several drawbacks. The parameter $k$ in~(\ref{eq:k_old1}) and~(\ref{eq:k_old2}) is chosen assuming that the load value is  known, which is not always the case in practical applications. Moreover, the computation of the parameter $k$ is done exploiting the knowledge of most of the system parameters. This may translate into poor robustness of the control algorithm since small variations of the actual value of the system parameters cause an erroneous computation of $k$, thus a wrong definition of the sliding manifold. 

An alternative approach has been proposed in~\cite{cavallo_energ18},  where possible values of the load have been preliminarily identified by using statistic approaches, but the robustness of the approach is still an open issue. It is clear that the preferred solution is the design of an adaptive control law, yielding more robust characteristics and the capability of achieving the control goal even when the system load is not known. In this work, the parameter $k(t)$ is chosen adaptively. In the case of  battery charging with constant current $\bar{x}_1$, $k$ is chosen so as to satisfy
\begin{equation} \label{eq:k_dot}
	\dot{k} = \gamma_1 (\bar{x}_1 - x_1)
\end{equation}
with $\gamma_1$ being a positive constant to be chosen and $\bar{x}_1$ the current reference to be tracked.
It will be shown that a first order Sliding Mode Control is enforced and the control law
\begin{equation}\label{eq:contr_BBCU}
	u = 
	\begin{cases}
	0 & \text{when $\sigma\leq0$} \\
	1 & \text{when $\sigma>0$}\\	
	\end{cases}       
\end{equation}
guarantees that the sliding manifold~(\ref{eq:slid_man}) is reached in finite time. Moreover, the closed-loop system converges to an asymptotically stable equilibrium point.

The above considerations are formalised  in the following Theorem. 
\begin{thm}[Adaptive Current Control]\label{thm:1}
Consider the system~(\ref{eq:buck_boost_model1})--(\ref{eq:buck_boost_model3}) and the control law~(\ref{eq:contr_BBCU}) where the sliding function is defined as in~(\ref{eq:slid_fun}) and the parameter $k$ is chosen adaptively according to~(\ref{eq:k_dot}). Assume 
\begin{eqnarray}
	&&x_{3u}^*/x_{2u}^* < 1 \label{eq:x2x3} \\
	&&x^*_{10} < \bar{x}_1 < x^*_{11}, \label{eq:Th1_hp5}\\	
	&&X_3^+<X_2^-\label{eq:Th1_hp6}
\end{eqnarray}
and let $|k|<K_{\max{}}$ with
\begin{equation} \label{eq:kmax_lim}
	K_{\max{}} < \min\left\lbrace \frac{X_3^-}{L|\psi_1|}, \frac{X_2^--X_3^+}{L\left(\frac{|\psi_2|}{C_H} + \psi_3 \right) }\right\rbrace  
\end{equation}
and $\psi_1 \coloneqq \min\left\lbrace \beta_H - \alpha X_2^+ , -\left(\beta_H - \alpha X_2^- \right) \right\rbrace $, $\psi_2 \coloneqq \min \left\lbrace X_1^-, - X_1^+\right\rbrace  $ and $\psi_3 \coloneqq \max\left\lbrace \beta_H - \alpha X_2^+ , -\left(\beta_H - \alpha X_2^- \right) \right\rbrace $.
Then, defining 
\begin{eqnarray} 
	a(\gamma_1) &=& \gamma_1 L{x_2^*}^3 - \frac{R_L}{4}\frac{{x_2^*}^3}{x_3^*} - \frac{1}{4}\frac{{x_3^*}^2}{\gamma_1 L{k^*}\bar{x}_1}\label{eq:aGamma}\\
	b(\gamma_1) &=& \alpha C_H - 3\gamma_1 L{k^*}\bar{x}_1 \label{eq:bGamma}
\end{eqnarray}
choosing $\gamma_1>0$ such that
\begin{equation}
	\gamma_1 < \min{} \left\lbrace \frac{X_3^-/L + K_{\max{}}\psi_1}{\left(X_1^+-\bar{x}_1 \right)X_2^+ }, \frac{1}{\left(\bar{x}_1-X_1^- \right)X_2^+ }\left[\frac{X_2^--X_3^+}{L} - K_{\max{}} \left( \frac{|\psi_2|}{C_H} + \psi_3\right) \right]  \right\rbrace 
\end{equation} 
and such  that 
\begin{equation}\label{eq:abGamma}
\nu=\min\{a(\gamma_1), b(\gamma_1)\}>0
\end{equation}
the closed-loop system has the property that for any $\epsilon>0$ there exist $T>0$ such that 
	\begin{equation}
		|\bar{x}_1 - x_1(t)|<\epsilon \qquad \forall t>T.
	\end{equation}
Moreover, the system state reaches the positively invariant set  $\mathcal S$~(\ref{eq:slid_man}) in finite time.  
\end{thm} 
\begin{pf}
The  proof is based on the Theory of Sliding Mode. 

First, the reaching and the existence conditions of the sliding mode are shown, then  the stability of the zero-dynamics in sliding regime is proven.	
In order to prove the reaching condition, we need to demonstrate that starting either from $\sigma(k(0),x(0)) < 0$ or from $\sigma(k(0),x(0)) > 0$, the sliding manifold $\sigma(k,x) = 0$ is reached in finite time. Let us consider the case $\sigma(k(0),x(0)) < 0$, thus $u \equiv 0$, first. In this case, the system exponentially tends towards $\left[ x_{10}^* \quad x_{20}^* \quad x_{30}^*\right]^T $ and, given (\ref{eq:k_dot}) and (\ref{eq:Th1_hp5}), $k$ eventually will be increasing. Therefore there must be a time $t^*$ such that the system trajectory crosses the sliding surface. Analogously, for $\sigma(k(0),x(0)) > 0$, the initial control input is $u \equiv 1$. In this case the system exponentially tends towards $\left[ x_{11}^* \quad x_{21}^* \quad x_{31}^*\right]^T $ and $k$ eventually decreases with time (note that, due to~(\ref{eq:ELEH}), $x^*_{11}>0$), hence, also in this case, the system trajectory crosses the sliding surface.
The sliding mode existence property is guaranteed by proving
\begin{equation}\label{eq:sigma_sigmad}
	\sigma \dot{\sigma} < -\omega|\sigma|,\quad \mathrm{as}\quad\sigma\to 0
\end{equation}	
with $\omega>0$. 
Basically, the proof is based on the computation of $\dot{\sigma}$ 
\begin{eqnarray}\label{eq:dot_sigma}
	\sigma\dot{\sigma}	&=& \left( \dot{k}x_2 + k\dot{x}_2 - \dot{x}_1\right) \sigma \\
	&=& \left[ - \frac{1}{2}\left(\frac{kx_1}{C_H} + \frac{x_2}{L}  \right) + \dot{k}x_2 + \frac{x_3}{L}  + k\left(\beta_H - \alpha x_2\right) \right]\sigma - \frac{1}{2}\left(\frac{kx_1}{C_H} + \frac{x_2}{L}  \right) |\sigma|, \\
	&=& \left( \phi_1-\phi_2 \right)\sigma - \phi_2|\sigma| \nonumber
\end{eqnarray}
with $\phi_1\coloneqq \dot{k}x_2 + \frac{x_3}{L}  + k\left(\beta_H - \alpha x_2\right)$ and $\phi_2 \coloneqq \frac{1}{2}\left(\frac{kx_1}{C_H} + \frac{x_2}{L}  \right)$.
Then it is easy to verify that $\sigma \dot{\sigma} < -\omega|\sigma|$ holds if and only if  it holds
\begin{equation}
	\begin{cases}
	\phi_1 > 0, \\
	2\phi_2 - \phi_1 > 0.
	\end{cases}
\end{equation}
Let us verify the first condition, namely $\phi_1> 0$ by taking into account the worst case scenario. It is easy to verify that $\phi_1 > 0$ holds if $\gamma_1$ is chosen such that
\begin{equation}
	\gamma_1 < \frac{X_3^-/L + K_{\max{}}\psi_1}{\left(X_1^+-\bar{x}_1 \right)X_2^+ }.
\end{equation}
Note that, in order for $\gamma_1$ to be positive it must hold
\begin{equation}
	K_{\max} < \frac{X_3^-}{L|\psi_1|}.
\end{equation}
Similarly, $\gamma_1$ can be properly selected in order to obtain $2\phi_2 - \phi_1 > 0$ which holds if
\begin{equation}
	\gamma_1 < \frac{1}{\left(\bar{x}_1-X_1^- \right)X_2^+ }\left[\frac{X_2^--X_3^+}{L} - K_{\max{}} \left( \frac{|\psi_2|}{C_H} + \psi_3\right) \right],
\end{equation}
where $K_{\max}$ must be chosen such that
\begin{equation}
	K_{\max} < \frac{X_2^--X_3^+}{L\left(\frac{|\psi_2|}{C_H} + \psi_3 \right)} 
\end{equation}
in order to guarantee positivity of $\gamma_1$. 
Thus the systems trajectory reaches the sliding manifold and remains onto it in finite time. Moreover, in order to consider the tightest estimate of the decay rate of $\sigma$, $\omega$ can be selected as 
\begin{equation}\label{eq:reach}
\omega = \min\left\lbrace \phi_1, 2\phi_2 - \phi_1\right\rbrace.
\end{equation}

Hereafter, stability of the system trajectories constrained on the manifold must be studied. Once on the manifold, the sliding function is identically zero, therefore $\sigma \equiv 0$ and $\dot{\sigma} = 0$. Solving $\dot{\sigma} = 0$ for $u$ with $\sigma \equiv 0$, the equivalent control can be computed
\begin{equation}\label{eq:ueq}
u_{eq}=\frac{LC_H}{(Lk(t)^2+C_H)x_{2}}\left[\left(\dot{k}(t) -\alpha k(t)\right)x_{2}+\frac{x_{3}}{L} +\beta_H k(t)\right]
\end{equation}
	Replacing~(\ref{eq:ueq}) in~(\ref{eq:buck_boost_model2}) and~(\ref{eq:buck_boost_model3}), the equations of the system sliding on the manifold are obtained
	\begin{eqnarray}
		\dot{k}&=&\gamma_1\left( \bar{x}_1 - kx_2 \right) \label{eq:reduced_syst1} \\
		\dot{x}_2&=&-\alpha x_2-\frac{1}{C_H}kx_2 u_{eq}+\beta_H\label{eq:reduced_syst2} \\
		\dot{x}_3&=&\!\frac{1}{C_L}kx_2-\frac{1}{R_L C_L}x_3+\beta_L\label{eq:reduced_syst3} 
	\end{eqnarray}
	Stability of the equilibrium point of this system can be studied resorting to Lyapunov theory. Preliminarily, the equilibrium point of the system is translated to the origin by using the new state $z\in\mathbf{R}^3$, with $z_1=k-k^*$, $z_2=x_2-x_2^*$ and $z_3=x_3-x_3^*$, the superscript $(\cdot)^*$ denoting the steady-state solution. Thus, in the new coordinates one has
	\begin{eqnarray}
		\dot{z}_1 &=& -\gamma_1 \left[ z_1 \left(z_2 + x_2^* \right) + {k^*}z_2 \right] 	\label{eq:translated_syst1} \\
		\dot{z}_2 &=& \frac{1}{L \left( z_1 + {k^*}\right)^2 + C_H }\left[-L\left( z_1 + {k^*}\right) \left( z_2 + x_2^* \right)\dot{z}_1 - \alpha C_Hz_2 - z_3 \left(z_1 + {k^*} \right)  - x_3^*z_1 \right] 																				\label{eq:translated_syst2} \\
		\dot{z}_3 &=& -\frac{1}{R_L C_L}z_3 - \frac{1}{C_L \gamma_1}\dot{z}_1						\label{eq:translated_syst3} 
	\end{eqnarray}
	Let us consider the Lyapunov function
	\begin{equation}
		V(z) = \frac{Lx_2^{*^2}}{2}z_1^2 + \frac{1}{2}z_2^2\left[L\left( z_1 + {k^*}\right)^2 + C_H  \right] + \frac{C_L}{2}z_3^2
	\end{equation}
	whose time derivative, after some computation, is
	\begin{equation}
		\dot{V}(z) = - z^T
			\begin{bmatrix}
				\gamma_1 L{x_2^*}^3  & 0 & 0  \\
				0      & \alpha C_H - \gamma_1 L{k^*}\bar{x}_1 & 0 \\
				0 & 0 & \displaystyle{\frac{1}{R_L}}  
			\end{bmatrix}z
		+ \gamma_1 L x_2^*z_1z_2^2 \left( z_1 + 2{k^*}\right) +z_1\left( x_2^*z_3 - x_3^*z_2 \right)		
	\end{equation}	 
After some algebraic manipulation, the following expression is obtained	 
	\begin{equation}
		\dot{V}(z) \leq -a(\gamma_1)z_1^2 -b(\gamma_1)z_2^2 -cz_3^2 + \frac{1}{2}\gamma_1 L x_2^*  \left( z_1^2 + z_2^2\right)^2
	\end{equation}
	where
\begin{equation} 
		c = \frac{1}{R_L} -\frac{1}{R_L}\frac{x_3^*}{x_2^*} 
	\end{equation}
In the above derivation the well-known inequality 
	\begin{equation}
	|xy| \leq \frac{1}{2}\left(\rho^2x^2+y^2/\rho^2 \right) 
	\end{equation}
	for any $\rho\in\mathbf{R}$ has been extensively used. 
	Note that $c>0$ due to~(\ref{eq:x2x3})$, a(\gamma_1)$ and $b(\gamma_1)$ impose respectively a lower and an upper bound on the choice of the gain $\gamma_1$. Considering hypothesis~(\ref{eq:abGamma}) and denoting $\hat{z} = [z_1 \quad z_2]$ 
	\begin{equation} \label{eq: lyap_dot}
		\dot{V}(z) \leq -\nu ||\hat{z}||^2 +  \frac{1}{2}\gamma_1 L x_2^*||\hat{z}||^2 - cz_3^2 
		\leq -||\hat{z}||^2 \left(\nu - \frac{1}{2}\gamma_1 L x_2^*||\hat{z}||^2 \right) - cz_3^2
	\end{equation}
	Therefore,~(\ref{eq: lyap_dot}) will be negative definite in a cylinder of radius $||\hat{z}||$ such that 
	\begin{equation}\label{eq:radius1}
		||\hat{z}|| < \sqrt{\frac{2}{\gamma_1 L x_2^*}}\nu
	\end{equation}
Hence local asymptotic stability of the origin and an estimate of the region of attraction have been proved. 
\end{pf}
Some remarks are now in order. 
\begin{rem}[Feasibility]\label{rem:feas}
Condition~(\ref{eq:x2x3}) simply says that there are well-defined LV and HV voltage side, in the sense that it is not possible that in some working condition HV becomes smaller then LV. Condition~(\ref{eq:Th1_hp6}) better quantifies this voltage relationship between LV and HV side. 
	Condition~(\ref{eq:Th1_hp5}) is obvious in practical applications, simply the current cannot exceed the values it has in the extreme cases ($u=0$ or $u=1$). 
Finally, condition~$|k|<K_{\max{}}$ can be easily ensured by using a saturated integral to compute $k$ from~(\ref{eq:k_dot})
\end{rem}
The second task of the control is to limit the generator current, and this is enforced by considering a reference value $\bar{x}_2$ that the HV capacitor voltage has to track (see also Remark~\ref{rem:impl}). Also in this case we consider the sliding manifold~(\ref{eq:slid_man}), with sliding function~(\ref{eq:slid_fun}). However, now the adaptation strategy changes as follows
\begin{equation} \label{eq:k_dot2}
	\dot{k} = \gamma_2 (x_2-\bar{x}_2)
\end{equation}
with $\gamma_2>0$.
Note that the reference voltage has to has to be selected so that 
\begin{equation}\label{eq:x_2ref}
0<\bar{x}_2<\frac{1}{2}\frac{R_D}{R_D+R_H}E_H\left[1+\sqrt{1+\frac{R_D+R_H}{R_D}\frac{C_H}{C_L}\left(\frac{E_L}{E_H}\right)^2}\right]
\end{equation}
where the upper bound comes from Corollary 2 in~\cite{cavallo:ijrnc17}.
Also in this case, the control law~(\ref{eq:contr_BBCU})
guarantees that the sliding manifold~(\ref{eq:slid_man}) is reached in finite time, as it will be shown below.
Stability of the adaptive law in the case of voltage control is considerably harder to show than in the case of current tracking. Preliminarily, define the following symbols
\begin{equation}\label{eq:DE}
\Delta E = \sqrt{E_L^2-4\bar{x}_2\frac{R_L}{R_H}\left[\left(1+\frac{R_H}{R_D}\right)\bar{x}_2-E_H\right]}-E_L
\end{equation}
\begin{equation}\label{eq:gam1}
\hat{\gamma}_{2}=\frac{2}{L}\frac{C_H}{C_L}\frac{1+\left(1+\frac{R_H}{R_D}\right)\frac{R_L}{R_H}\frac{C_L}{C_H}}{\Delta E}+\frac{\Delta E}{2C_LR_L^2\bar{x}_2^2}
\end{equation}
\begin{eqnarray}
a_{0} &=& \frac{1}{2}\left(\Delta E+E_L\right)\\
a_{10}&=&\frac{1}{\bar{x}_2}\left(\frac{E_H}{R_H}-\frac{E_L}{R_L}\frac{\Delta E-E_L}{2\bar{x}_2}\right)\\
a_{11}&=&\left[\frac{L}{R}\left(\Delta E-E_L\right)+R_LC_L\left(\Delta E+E_L\right)\right]\\
a_{20}&=&\frac{1}{2}LC_L\hat{\gamma}_{2}\left(\Delta E-E_L\right)\\
a_{21}&=&\frac{1}{2}LC_L\left(\Delta E-E_L\right)\\
a_3&=&\left(R_LC_L\right)^2\left[\frac{1}{2}\frac{L}{R_L}\hat{\gamma}_{2}\left(\Delta E -E_L\right)-\frac{R_D+R_H}{R_DR_H}\right]
\end{eqnarray}
Finally, define the polynomial equation
\begin{equation}\label{eq:pol_gamma}
p(\gamma)=a_{11}a_{21}\gamma^2+\left(a_{11}a_{20}+a_{10}a_{21}-a_3a_0\right)\gamma+a_{10}a_{20}=0
\end{equation}
and let  $\gamma^+$ be the smallest  {\em real positive} solution (if any) of~(\ref{eq:pol_gamma}). If~(\ref{eq:pol_gamma}) has only complex or real negative solutions, we let $\gamma^+=\infty$.  Now we are in the position to state the following Theorem.
\begin{thm}[Stability on the manifold, Voltage Control]\label{thm:2}
Consider the system~(\ref{eq:buck_boost_model1})--(\ref{eq:buck_boost_model3}) and the control law~(\ref{eq:contr_BBCU}) where the sliding function is defined as in~(\ref{eq:slid_fun}) and the parameter $k$ is chosen adaptively according to~(\ref{eq:k_dot2}). Suppose 
\begin{equation}\label{eq:RDhigh}
R_D >\frac{\bar{x}_2}{E_H-\bar{x}_2}R_H
\end{equation} 
Then $\gamma_2$ can be any positive scalar. If otherwise 
\begin{equation}\label{eq:RDlow}
R_D \le\frac{\bar{x}_2}{E_H-\bar{x}_2}R_H
\end{equation} 
then select $\gamma_2<\min(\hat{\gamma}_{2},\gamma^+)$. Assume
\begin{eqnarray}
&&x^*_{21} < \bar{x}_2 < x^*_{20}, \label{eq:Th2_hp1}\\	
&&X_3^+<X_2^-\label{eq:Th2_hp2}\\
\end{eqnarray}
and let $|k|<K_{\max}$ with $K_{\max}$ as in~(\ref{eq:kmax_lim}) and $\psi_1$, $\psi_2$ and $\psi_3$ selected as in Theorem~\ref{thm:1}.
Then, choosing $\gamma_2>0$ such that the additional condition
\begin{equation}
\gamma_2 < \min{} \left\lbrace \frac{X_3^-/L + K_{\max{}}\psi_1}{\left(\bar{x}_2-X_2^-\right)X_2^+ }, \frac{1}{\left(\bar{x}_2-X_2^-\right)X_2^+ }\left[\frac{X_2^--X_3^+}{L} - K_{\max{}} \left( \frac{|\psi_2|}{C_H} + \psi_3\right) \right]  \right\rbrace, 
\end{equation} 
is satisfied, the closed-loop system converges locally asymptotically to a unique steady-state solution with $x_2^*=\bar{x}_2$, and for any $\delta>0$ there exists a $T>0$ such that
	\begin{equation}
		|\bar{x}_2 - x_2(t)|<\delta \qquad \forall t>T
	\end{equation}
Moreover, the system state reaches the positively invariant set  $\mathcal S$~(\ref{eq:slid_man}) in finite time.  
\end{thm} 
\begin{pf}
As in the proof of Theorem~\ref{thm:1}, we have to prove that the sliding manifold is reached first, and then the stability of the system on the manifold. The proof of reaching follows the same steps as in Theorem~\ref{thm:1}, with the due modification of the upper bound on $\gamma_2$, and is omitted for the sake of brevity. Once the sliding has been established, The stability on the manifold has to be proved. The equivalent control is again~(\ref{eq:ueq}), but~(\ref{eq:reduced_syst1}) changes to~(\ref{eq:k_dot2}). 
Consider the coordinate translation 
\begin{eqnarray}
z_1&=&k-k^*\\
z_2&=&x_2-\bar{x}_2\\
z_3&=&x_3-x_3^*
\end{eqnarray}
where $k^*=x_1^*/\bar{x}_2$. Then the dynamic of the translated system is given by
\begin{eqnarray}
\dot{z}_1 &=& \gamma_2 z_2\label{eq:translated_syst12} \\
\dot{z}_2 &=& \frac{1}{L \left( z_1 + {k^*}\right)^2 + C_H }\left[-L\left( z_1 + {k^*}\right) \left( z_2 + \bar{x}_2 \right)\gamma_2 z_2 - \alpha C_H z_2 - z_3 \left(z_1 + {k^*} \right)  - x_3^*z_1 \right]\label{eq:translated_syst22} \\
\dot{z}_3 &=& -\frac{1}{R_L C_L}z_3 + \frac{1}{C_L }\left(z_1z_2+k^* z_2+\bar{x}_2 z_1\right)\label{eq:translated_syst32} 
\end{eqnarray}
Local stability of the origin of the system~(\ref{eq:translated_syst12})--(\ref{eq:translated_syst32}) can be assessed locally by linearization. The dynamic matrix can be computed as
\begin{equation}\label{eq:dyn_mat2}
A=\left[
\begin{array}{ccc}
0&\gamma_2&0\\
-\frac{x_3^*}{L{k^*}^2+C_H}&-\frac{\alpha C_H+\gamma_2Lk^*\bar{x}_2}{L{k^*}^2+C_H}&-\frac{k^*}{L{k^*}^2+C_H}\\
\frac{\bar{x}_2}{C_L}&\frac{k^*}{C_L}&-\frac{1}{R_LC_L}
\end{array}
\right]
\end{equation}
whose characteristic polynomial is 
\begin{eqnarray}\label{eq:poly}
q(s)&=&s\left(s^2+\frac{L{k^*}^2+C_H+\alpha C_HR_LC_L}{R_LC_L(L{k^*}^2+C_H)}s+\frac{R_L{k^*}^2+\alpha C_H}{R_LC_L(L{k^*}^2+C_H)}\right)\nonumber\\
&+&\gamma_2\left(\frac{Lk^*\bar{x}_2}{L{k^*}^2+C_H}s^2+\frac{R_LC_Lx_3^*+Lk^*\bar{x}_2}{L{k^*}^2+C_H}s+\frac{R_Lk^*\bar{x}_2+x_3^*}{L{k^*}^2+C_H}\right)\triangleq a(s)+\gamma_2b(s)\label{eq:p}
\end{eqnarray}
Stability of system~(\ref{eq:translated_syst12})--(\ref{eq:translated_syst32}) for different values of $\gamma_2$ can be analyzed by using root locus analysis. Note that, although $\bar{x}_2>0$, the sign of $k^*$ is not known {\em apriori}. 
However, if~(\ref{eq:RDhigh}) holds, it is possible to show, by algebraic calculation, that $k^*>0$. In this case both the polynomials $a(s)$ and $b(s)$ in~(\ref{eq:p}) have roots in the closed complex left half-plane, and the stability is assured by any $\gamma_2>0$.  On the contrary, if~(\ref{eq:RDlow}) holds, $k^*$ is surely negative. In this case the stability holds for $\gamma_2>0$ only if $2R_Lk^*\bar{x}_2+E_L$ is imposed positive. In this case the polynomial $b(s)$ has one positive and one negative root, hence there will be an upper bound to the stabilizing values of $\gamma_2$. The upper bound can be sought by using Routh-Hurwitz criterion. After some algebraic computation, the following stability conditions must be satisfied.
\begin{equation}
\gamma_2<\hat{\gamma}_2
\end{equation}
and, if the polynomial $p(\lambda_2)$ has real roots, denote them by $\gamma_2'$ and $\gamma_2''$, with $\gamma_2'<\gamma_2''$, 
\begin{eqnarray}
\gamma_2<\gamma_2''&&{\rm if\  } a_{11}a_{21}<0\\
\gamma_2<\gamma_2' {\rm \ or\ } \gamma_2>\gamma_2''&&{\rm otherwise}
\end{eqnarray}
note that if $a_{11}a_{21}<0$, then $\gamma_2'<0$. 
The above conditions are implied by $\gamma_2<\min(\hat{\gamma}_{2},\gamma^+)$. This concludes the proof.
\end{pf}

\begin{rem}[Adaptation gain]
From the proof of Theorem~\ref{thm:2} it can be shown that if the load satisfies~(\ref{eq:RDlow}), then a {\em negative} $\gamma_2$ could be sought to stabilize the closed-loop system. However, in practical applications often the load is unknown, and since negative $\gamma_2$ are destabilizing for the case (\ref{eq:RDhigh}), the solution is to use the smallest {\em positive} $\gamma_2$ stabilising any load. 
\end{rem}
\begin{rem}[Reaching time]
It is easy to estimate the reaching 
time~\cite[Chap.~7]{slotine1991applied} for the case of battery current regulation
\begin{equation}\label{eq:reaching}
t_{reach}\le \sigma(0)/\omega
\end{equation}
with $\omega$ given in~(\ref{eq:reach}). With the same approach, a similar estimate holds also for the control strategy presented in Theorem~\ref{thm:2}. Details are omitted. 
\end{rem}

\begin{rem}[Robust implementation]\label{rem:impl}
The choice of the adaptation law~(\ref{eq:k_dot2}) is based on the algebraic relationship 
$x_2=E_H-R_HI_{g}$
so that, the generator current can be limited to a prescribed overload current $I_{OL}$ by considering a reference voltage 
\begin{equation}\label{eq:x2ref}
\bar{x}_2=E_H-R_HI_{OL}.
\end{equation}
However, the estimate of $\bar{x}_2$ is prone to errors due to possible uncertainties in $R_H$ and $E_H$. Hence, assuming the  generator current measurement available, a better and more robust adaptive law is simply
\begin{equation}\label{eq:k_dot2i}
\dot{k}=\gamma_2'(I_{OL}-I_{g})
\end{equation}
where $\gamma_2'=R_H\gamma_2$ is  a positive gain.
\end{rem}

\begin{rem}[ROA estimate]\label{rem:ROA}
From the above derivation, it is clear that obtaining an analytic estimate of the Region of Attraction (ROA) is considerably harder in the case of voltage control, Theorem~\ref{thm:2} than in the case of Theorem~\ref{thm:1}. However, a numeric estimate of the ROA is possible by using a techniques proposed in~\cite{chesi2011} and starting from a Lyapunov function computed for the linearised case. 
In particular, since we are interested in reaching the steady-state within $5$s, for a given load $R_D$ and reference voltage $\bar{x}_2$, considering the model~(\ref{eq:translated_syst12})--(\ref{eq:translated_syst32}) compute the dynamic matrix $A$~(\ref{eq:dyn_mat2})  such that the related Lyapunov function $V=x^TPx$ has decay rate less than $0.75$ (so that after about $3$s the transient can be considered vanished). This can be accomplished   by  solving the Lyapunov equation 
\begin{equation}\label{eq:lyap}
(A^T+0.75I_3)P+P(A+0.75I_3)=-I_3
\end{equation}
for a positive solution $P$, 
being $I_3$ the $3\times 3$ identity matrix. 
Then define the Lyapunov function $V(z)=z^TPz$, with $z=(z_1,z_2,z_3)^T$. 
Note that, being $V(z)$ quadratic,  its derivative along the trajectory of~(\ref{eq:translated_syst12})--(\ref{eq:translated_syst32}), $\dot{V}=\Delta V \dot{z}$ is a fractional function, since the gradient $\Delta V$ is linear in $z$ and the entries of $\dot{z}$ are ratios of polynomial in $z$. Thus, it is easy to show that $\dot{V}$ is the ratio of a rather complex polynomial numerator, call it $N(z)$ and the {\em positive} polynomial $L(z_1+k^*)^2+C_H$. Thus, the problem of computing an estimate of the ROA can be reduced to solving a sequence of SDP~\cite[Chapter 2.2]{chesi2011}. Note that in general the positive solution of~(\ref{eq:lyap}) is not guaranteed to exist. Another possibility is to start from the Lyapunov function {\em maximising} the decay rate of the closed-loop system~\cite{LMI:94}, obtained by solving the GEVP
\begin{eqnarray}
&\max{}\lambda\qquad \nonumber & \\
&{\mathrm{subject \ to}} &P>0,\quad A^TP+PA+2\lambda P\le 0\label{eq:min_dec}
\end{eqnarray}
and then again using the procedure in~\cite{chesi2011} for the estimate of the ROA. 
The effectiveness of this approach will be illustrated in Section~\ref{sect:simul}.  
\end{rem}
\subsection{High-Level Control}\label{sec:supervisor}
In view of what has been expressed above, it is clear that the entire BBCU has at least two main operational modalities: charging the battery, when the generator can accomplish the objective of feeding the grid loads keeping its current below a prescribed threshold, and regulating the generator current to a prescribed level in order to let the battery help the generator when an overload occurs. This can be accomplished with a simple automaton using just two modes. Specifically

\begin{itemize}
	\item Mode 1: the generator on the high voltage side recharges the battery with a constant current $\bar{x}_1$ choosing the adaptive parameter $k(t)$ as in~(\ref{eq:k_dot}).
	\item Mode 2: if an overload occurs the supervisor must commute to Mode 2 to regulate the high voltage capacitor voltage (equivalently, the generator current) to a prescribed voltage set-point $\bar{x}_2$ (equivalently, $I_{OL}$), possibly asking the battery to provide energy to loads. In this case, the the adaptive parameter $k(t)$ is chosen as in as in~(\ref{eq:k_dot2i}).
\end{itemize}
It must be noted that, in order to avoid fast switching between modalities, an hysteresis with band $\left[ I_{OL} - \eta,  I_{OL} + \eta\right] $ is used rather than a strict threshold. \\

Moreover, for ensuring a safe commutation between the two Modes, an estimate of the region of attraction is computed in both operational modes. The overall strategy works as follows. Initially the system is in Mode 1 (assumed in the region of attraction of the controller). 
Next, if an overload occurs, Mode 2 is activated. However, before entering Mode 2 we must be sure that, after the finite  time needed to reach the new sliding manifold, the system state belongs to the region of attraction of the new active controller configuration, i.e., Mode 2. This is done comparing the current state and the estimate of the region of attraction for the new controller configuration. If the state is in the interior of the region of attraction, then the transition to Mode 2 is enabled, otherwise, a reduced performance mode is activated, with $I_{OL}$ increased so that the current state is within the reduced-performance ROA. This point will be clarified in Section~\ref{sect:large}. Next, the generator current is reduced, by slowly reducing the $I_{OL}$ until the original performance are achieved. A similar idea can be found in~\cite{CavalloAutomatica17b}.

%

\section{Numerical estimate of the ROA and simulation results}\label{sect:simul}
\begin{figure}[htb]
	\centering
	\includegraphics[width=0.9\linewidth]{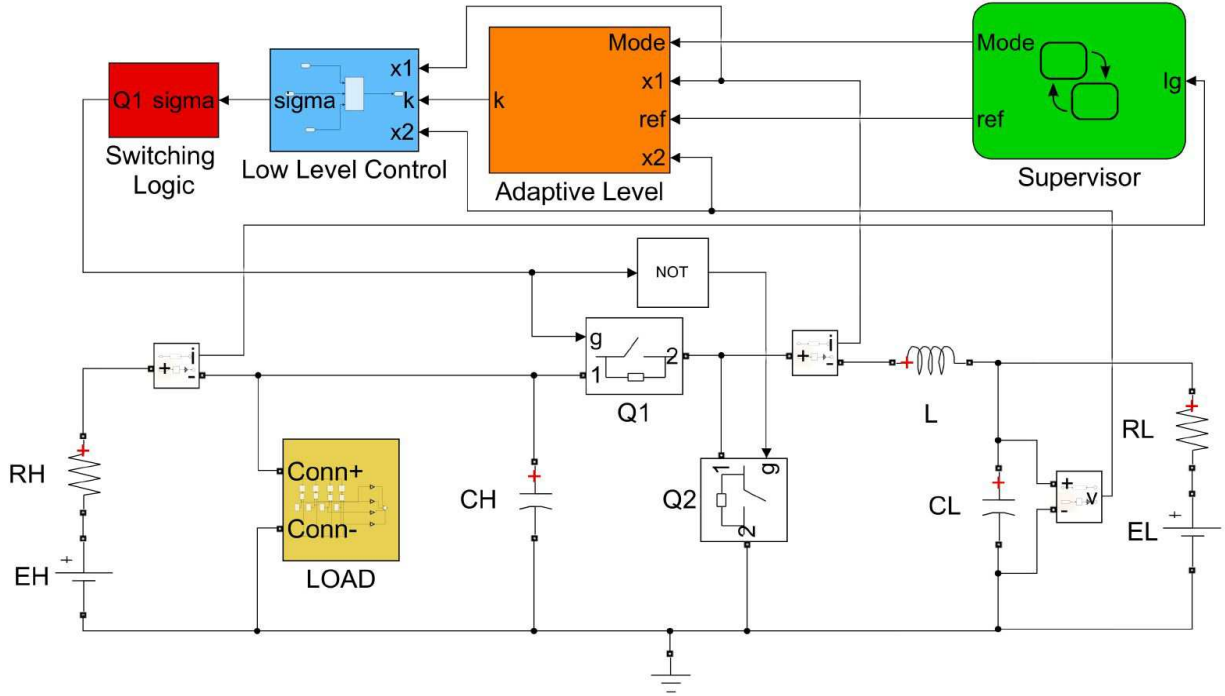}
	\caption{Simulink scheme of controlled system}
	\label{fig:schemesim}
\end{figure}

The proposed adaptive sliding control and associated supervisory strategy for MEA have been tested in a detailed MATLAB/Simulink/SimPowerSystem simulator, shown in Figure~\ref{fig:schemesim}, and composed of five blocks:

\begin{itemize}
	\item LOAD: contains a bank of parallel resistors (yellow block).
	\item SUPERVISOR: implements the High-Level Control designed according to Section~\ref{sec:supervisor} (green block).
	\item ADAPTIVE LEVEL: estimates the parameter $k$ according to~(\ref{eq:k_dot}) (orange block).
	\item LOW LEVEL CONTROL: implements the Low-Level Control logic according to Section~\ref{sec:lowlevel} (light blue block).
	\item SWITCHING LOGIC: realizes the switching logic for $Q_1$ and $Q_2$ switches (red block).
\end{itemize}


\begin{figure}[htb]
	\centering
	\includegraphics[width=0.6\linewidth]{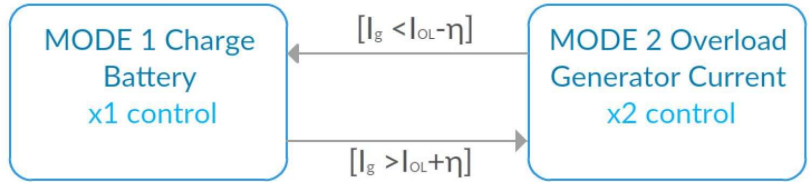}
	\caption{Supervisor}
	\label{fig:supervisorysim}
\end{figure}

The supervisor is a simple two-modes automaton, as shown in Figure~\ref{fig:supervisorysim}, and it is been implemented using the MATLAB StateFlow toolbox.

The system and controller parameters are shown in Tables~\ref{tab:bbcu_data}-\ref{tab:contr_par}. 
The reaching time is estimated according to~(\ref{eq:reaching}) for the inductor current control, and is $0.14$s. A similar computation for the generator current control gives an estimate of the reaching time for the second sliding surface of $0.02$s, that will be neglected in the following discussion, due to its small value. Two scenarios have been simulated, as presented in the following sections. 

\begin{table}[htb] 	
	\begin{subtable}{.5\linewidth} 
		\centering{
			\begin{tabular}{lll}
				\toprule
				Parameter & Value &  \\
				\midrule
				$E_H$ & $ 270 $ &  [V]  \\
				\midrule
				$R_H$& $ 100 $ &[m$\Omega$]  \\
				\midrule
				$L$ & $ 10 $ & [mH]  \\
				\midrule
				$C_{H}$& $ 0.8 $& [mF]\\
				\midrule
				$E_L$ & 28 &[V] \\ 
				\midrule
				$R_L$ & 100 & [m$\Omega$] \\ 
				\midrule
				$C_{L}$& $ 0.4 $ &[mF]\\ 
				\bottomrule
		\end{tabular}}           
		\caption{System Parameters}\label{tab:bbcu_data}
	\end{subtable}%
	\begin{subtable}{.5\linewidth}
		\centering{			
			\begin{tabular}{lll}
				\toprule
				Parameter &  Value&\\ 
				\midrule
				$\gamma_1$ &$ 4 $& [V$\cdot$s]$^{-1}$\\ 
				\midrule
				$\gamma_2$ &$ 4 $& [V$\cdot\Omega\cdot$s]$^{-1}$\\ 
				\midrule
				$\bar{x}_1$ &$ 10 $& [A]\\ 
				\midrule
				$I_{OL}$& $16$ & [A]\\ 
				\midrule
				$\eta$ &$ 0.5 $& [A]\\ 
				\bottomrule
		\end{tabular}}
		\caption{Controller Parameters} \label{tab:contr_par}
	\end{subtable}\caption{Simulation parameters}
\end{table}

\subsection{Scenario 1: small load variations}\label{sec:sim1}
\begin{figure}[htb]
	\centering
	\includegraphics[width=0.5\linewidth]{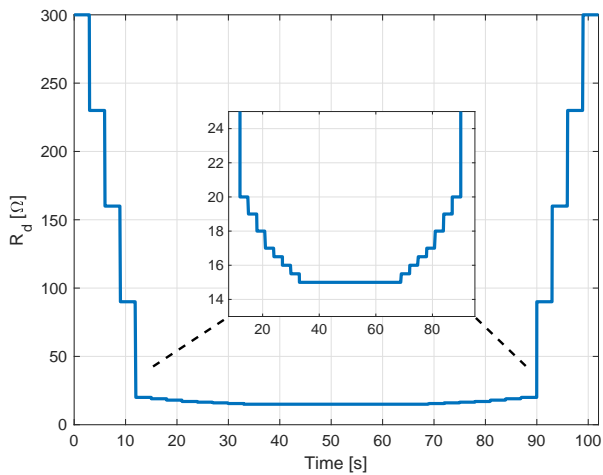}
	\caption{Load variation in Scenario 1}
	\label{fig:load_s1}
\end{figure}

A preliminary set of simulations has been carried out by considering a stepwise constant load $R_D$ varying from $300\Omega$ to $15\Omega$. Since, as shown by~(\ref{eq:k_old1}), (\ref{eq:k_old2}), for low values of $R_D$ the parameter $k$ varies widely,  then nonuniform variation of $R_D$ has been considered. 
Specifically, in the interval $[20,300]\Omega$ a large step of $70\Omega$ has been selected, in $[17,20]\Omega$ interval the step has been reduced to $1\Omega$. Finally, the interval $[15,17]\Omega$ has been swept with step interval $0.5\Omega$. 
The varying load is shown in Figure~\ref{fig:load_s1}, with a zoom around the zone with smaller variation of $R_D$. 

\begin{figure}[htb]
	\centering
	\includegraphics[width=0.5\linewidth]{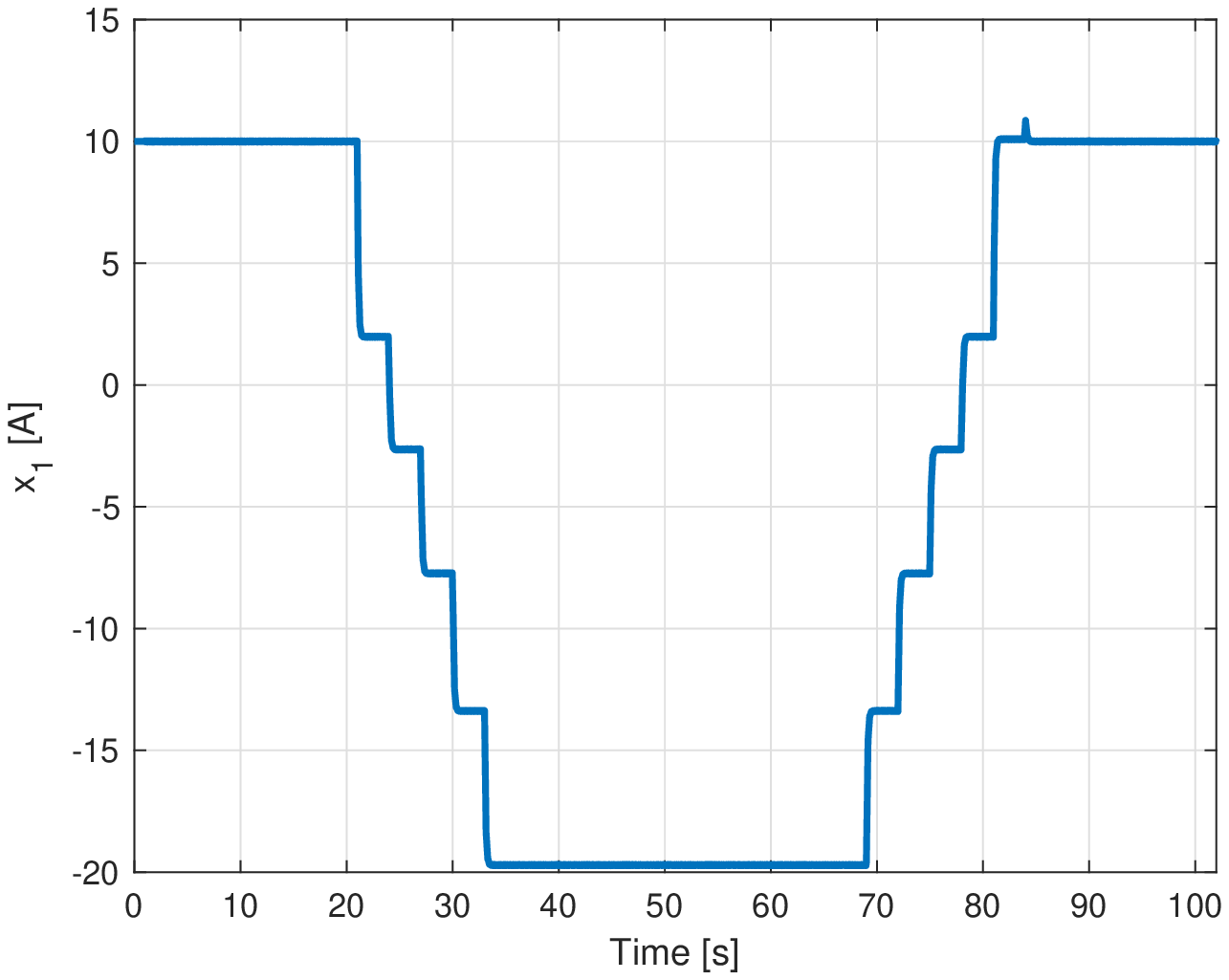}
	\caption{Inductor current, Scenario 1}
	\label{fig:x1s1}
\end{figure}

The purpose of this first set of simulations is to gain insight in the variation of the gains $k$ associated to the loads and an estimate of the ROA's for the controlled plant with a reasonable set of loads. 
In this first set of simulations the overload current is fixed to $I{OL}=16$A. 
The simulations starts with $R_{D}=300\Omega$. The generator current is below the prescribed threshold ($I_{OL}$), therefore no overload occurs. The supervisor is initially in Mode $1$, hence the objective is to recharge the battery with constant current ($\bar{x}_1$) through the inductor, as shown in Figure~\ref{fig:x1s1} (first $21$s). 

\begin{figure}[htb]
	\centering
	\includegraphics[width=0.5\linewidth]{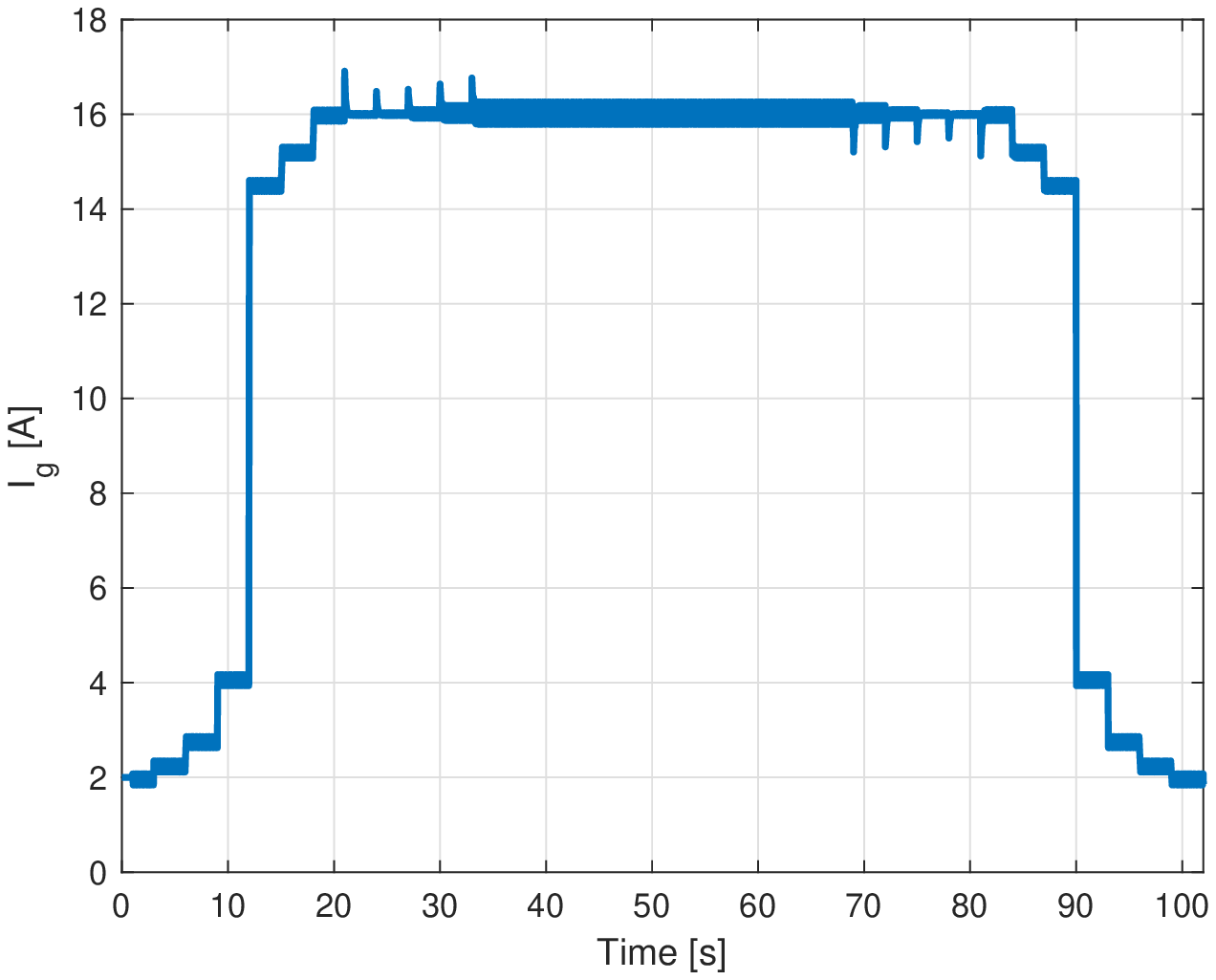}
	\caption{Generator current, Scenario 1}
	\label{fig:Igens1}
\end{figure}

Every $3$s the load resistor is decreased, and the supervisor for the first $18$s remains in Mode $1$, with generator current increasing (Figure~\ref{fig:Igens1}), but without reaching the threshold of maximum current. 
At time $t=18$s, the load becomes $R_D=18\Omega$ and the current request to the generator becomes $I_g=16.4$A. Note that, although the requested current exceeds $I_{OL}$, the value $I_{OL}+\eta$ is {\em not} exceeded, hence the supervisor remains in Mode $1$. After 3 more seconds, at $t=21$s, the load $R_D=17\Omega$ makes the generator current  exceed $I_{OL}+\eta$, hence the supervisor switches to Mode $2$, changing the control objective to drive the generator current to the current threshold $I_{OL}$. 

In order to guarantee the stability, some considerations on the ROA's are in order. 
When in  Mode $1$, the ROA is the cylinder with radius~(\ref{eq:radius1}). Using the numerical values above and Theorem~\ref{thm:1}, we have a rough estimate of the ROA is $||\hat{z}||< 4.3$, that is a large region in our case. Thus, in this case study, we can safely assume that any load variation in Mode $1$ happens when the controlled state is {\em within} the ROA. 

\begin{figure}[htb]
	\centering
	\includegraphics[width=0.5\linewidth]{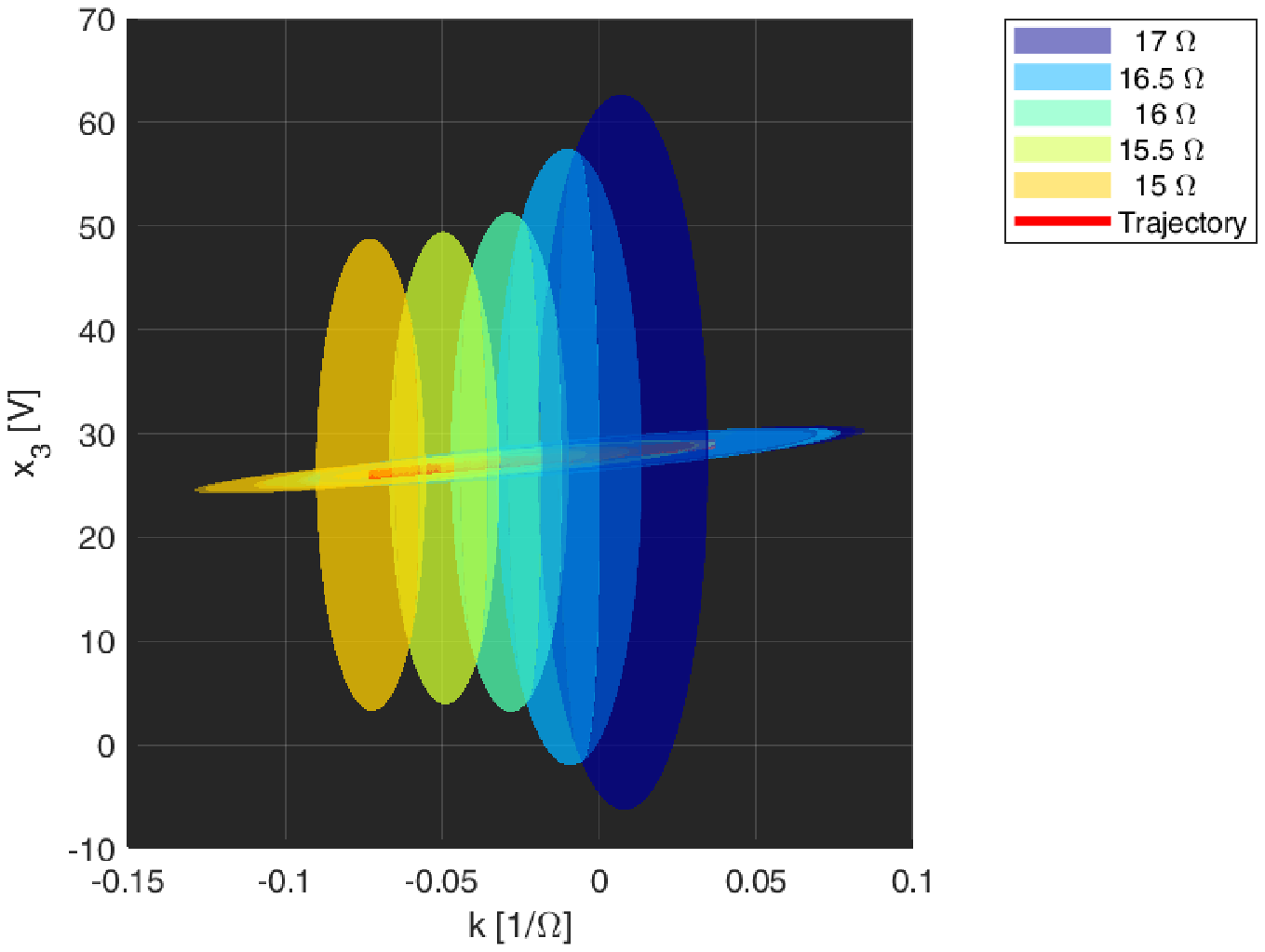}
	\caption{Variation of the Domain of Attraction as the load varies smoothy}
	\label{fig:roa_Ig_Rd}
\end{figure}

The situation is different when in Mode $2$, where, moreover, larger variations of the variables happen. In Figure~\ref{fig:roa_Ig_Rd} the ROA's are shown for different loads. Although the ROA is a 3D region, only its projection on the $(x_3,k)$ plane will be presented, for the sake of clearness. Two estimates of the ROA are computed for each load: one resulting from~(\ref{eq:lyap}) (thick region) and one from~(\ref{eq:min_dec}) (thin region). The union of the two regions is in the ROA. 
Note that at $t=21$s, when the supervisor decides the commutation from Mode 1 to Mode 2, the large blue ROA, that is the ROA of the new control action,  includes the current state, hence the supervisor can safely switch to the strategy in Theorem~\ref{thm:2}.
The remaining load variation cause related changes in the steady-state of the closed-loop system. Figure~\ref{fig:roa_Ig_Rd} shows that if the change of load is small and slow enough, stability is preserved. Indeed, for the values of $R_D$  shown in Figure~\ref{fig:roa_Ig_Rd}, note that the centre of the ROA associated to a given load belongs to the interior of the ROA associated to the next load (assuming the sequence of loads defined by the load profile in Figure~\ref{fig:load_s1}): e.g., the centre of the blue region is within the elongated azure region pertaining to $R_D=16.5\Omega$, and the centre of the azure region is within the elongated region of the cyan region related to $R_D=16\Omega$. This shows that a smooth and ``slow'' transition from $R_D=17\Omega$ to $R_D=16\Omega$ preserves stability. Note that ``slowness'' is related to the optimal decay rate computed by means of~(\ref{eq:min_dec}), hence it can be precisely estimated.  In Figure~\ref{fig:roa_Ig_Rd} also the actual trajectory of the system (in red) is shown. Note that, since the ROA's are positively invariant sets, no trajectory exits its ROA. 

\subsection{Large loads variations}\label{sect:large}
\begin{figure}[htb]
	\centering
	\includegraphics[width=0.5\linewidth]{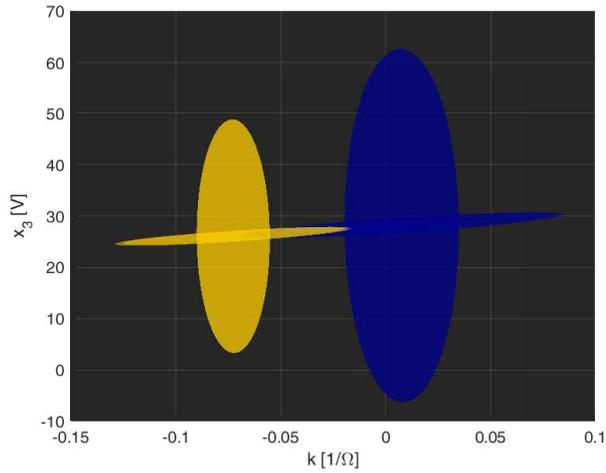}
	\caption{Variation of the Domain of Attraction as the load varies abruptly. Blue: $R_D=17\Omega$; yellow: $R_D=15\Omega$}
	\label{fig:roa_IOL_Rd}
\end{figure}

Different is the case where there is an abrupt change of load. For instance, in Figure~\ref{fig:roa_IOL_Rd} it is clear that, since the centre of the blue region does not intersect the yellow region, there is no guarantee that the transition from $R_D=17\Omega$ to $R_D=15\Omega$ will preserve stability. 

A possible solution to this issue is adopt a ``transient reduced performance'' approach, i.e.,  to temporarily relax the constraints. If we increase the   overload current, we have that, with the same $R_D$, the steady-state of $k$ increases (since less current is needed from the battery) and this observation ban be applied to move the leftmost region in Figure~\ref{fig:roa_IOL_Rd} to the right. 

\begin{figure}[htb]
	\centering
	\includegraphics[width=0.5\linewidth]{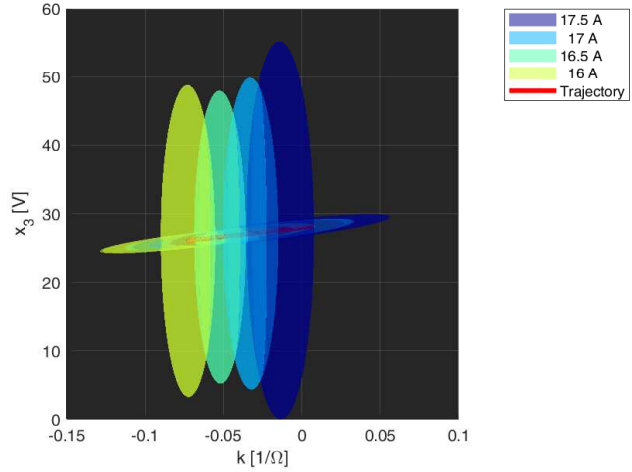}
	\caption{Variation of the Domain of Attraction as  $I_{OL}$ varies with $R_D=15\Omega$. }
	\label{fig:roa_Rd_IOL}
\end{figure}

To gain insight in the above consideration, we fix the load to $R_D=15\Omega$ and compute the estimate of the ROA by considering different overload current  assuming values in the interval $[16,17.5]$A and varying with steps of $0.5$A. The result is depicted in  Figure~\ref{fig:roa_Rd_IOL}, that suggests us how to guarantee stability when the load changes abruptly. Simply, when the load goes to $R_D=15\Omega$, imposing $I_{OL}=17.5$A makes the ROA to include the centre of the ROA with $R_D=17\Omega$, $I_{OL}=16$A, thus preserving stability. Next, by changing the reference smoothly to $I_{OL}=16$A the original performance are restored with assured stability. 

A final consideration is in order. The selection of a reduced performance overload current depends on the knowledge of an estimate of the load. In practical applications sometimes even a rough estimate of the load is not available. In this case a viable strategy is to assume the worst-case scenario, i.e., large variations of the load, and use the ``transient reduced performance'' approach. 

\subsection{Scenario 2}
\begin{figure}[htb]
	\centering
	\includegraphics[width=0.5\linewidth]{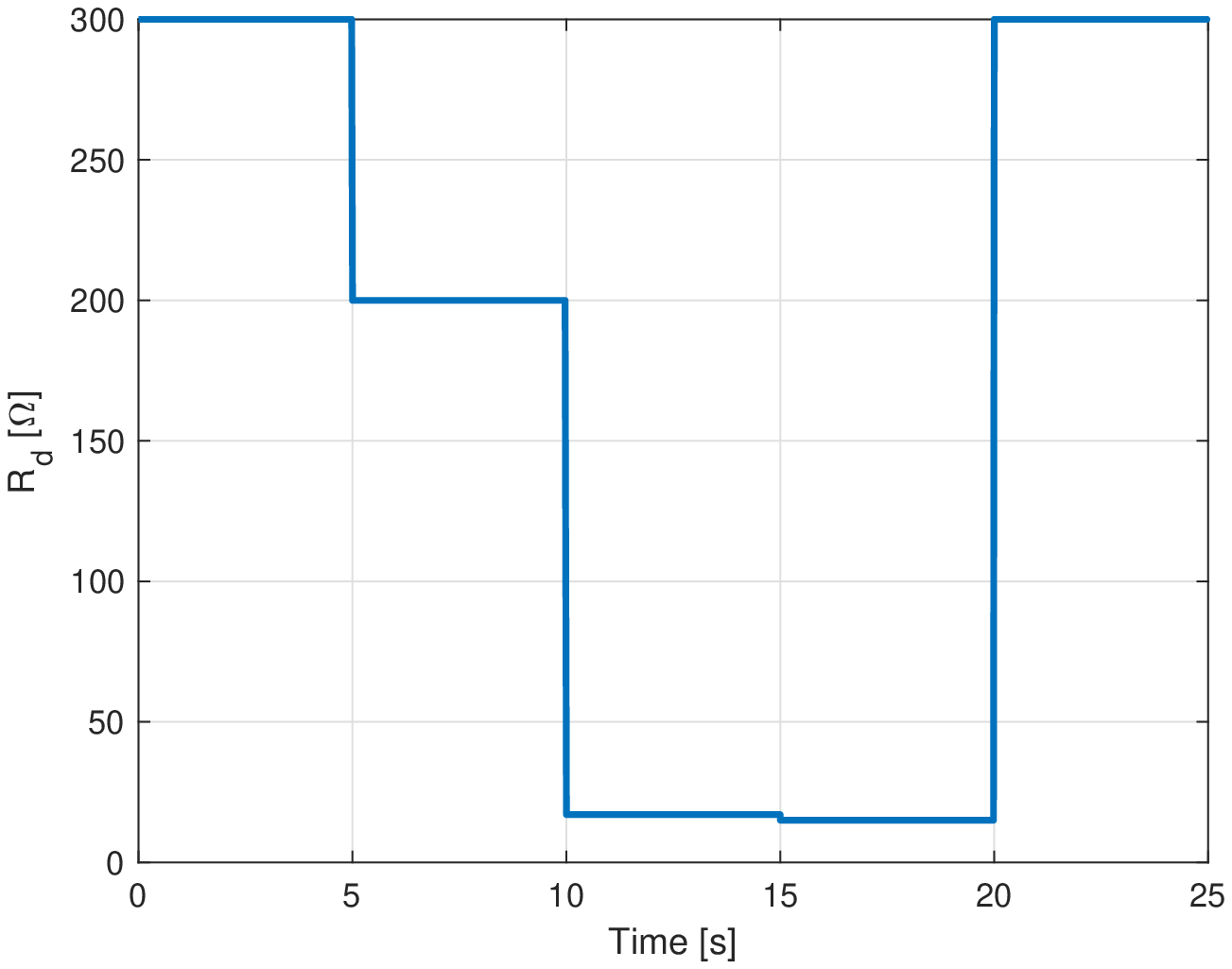}
	\caption{Load variation in Scenario 2}
	\label{fig:load_s2}
\end{figure}

The second set of simulations has been carried out referring to the large and unknown load variations in Figure~\ref{fig:load_s2} and Table~\ref{tab:load_s2}.

\begin{table}[htb] 
	
	\centering
		\begin{tabular}{lll}
			\toprule
			Parameter & Resistance [$\Omega$] & Activation Time [s]  \\
			\midrule
			$R_{D1}$ &$ 300 $& $[0-5] [20-25]$  \\
			\midrule
			$R_{D2}$ &$ 200 $& $[5-10]$  \\
			\midrule
			$R_{D3}$& $17$ & $[10-15]$  \\
			\midrule
			$R_{D4}$ &$ 15 $& $[15-20]$\\
			\bottomrule
	\end{tabular}\caption{Load variation}\label{tab:load_s2}
\end{table}

In this case the reference generator current $I_{OL}$ is step-wise varied  in  time in order to guarantee  stability,  so that the initial condition belongs  to the ROA. The reference is varied after it has approximately reached its steady-state. The time required for this  is estimated based on the worst case of decay time estimated when solving the GEVP~(\ref{eq:min_dec}). Using the above numerical values, one can assess that in the worst case, at least $0.79$s are required to reach the steady state with $90\%$ accuracy. 

\begin{figure}[H]
	\centering
	\includegraphics[width=0.5\linewidth]{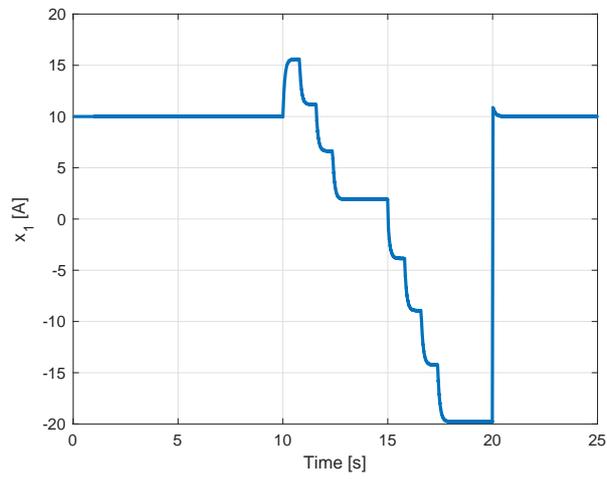}
	\caption{Inductor current, Scenario 2}
	\label{fig:x1s2}
\end{figure}

\begin{figure}[H]
	\centering
	\includegraphics[width=0.5\linewidth]{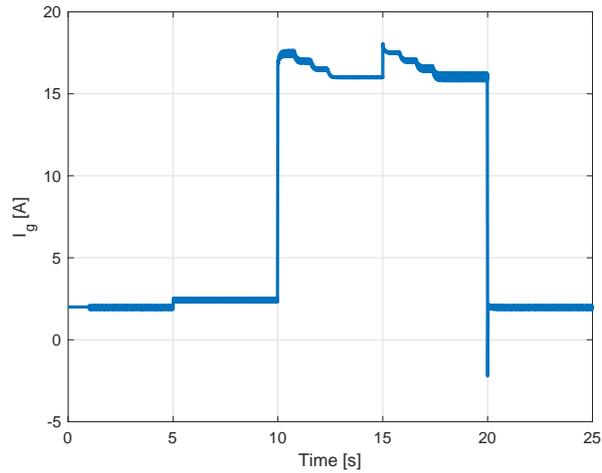}
	\caption{Generator current, Scenario 2}
	\label{fig:Igens2}
\end{figure}

At the beginning of the simulation, the supervisor is in Mode $1$ with load $R_D = R_{D1} = 300 \Omega$ so the low-level goal is to control the inductor current $x_1$ to $\bar{x}_1$ = $10$A  as shown in Figure~\ref{fig:x1s2}. 
Five seconds later a new load is added, and $R_D$ becomes $R_{D2}$. In this step the generator current increases (Figure~\ref{fig:Igens2}), but it is again below  the threshold of maximum current hence the supervisor remains in Mode $1$. As stated in Section~\ref{sec:sim1}, load variations always leave the state of the controlled system in the ROA of the current objective, thus stability in Mode $1$ is ensured. At time  $t=10$s an additional load is inserted, producing a power request  exceeding overload. The supervisor reacts by switching to Mode $2$, i.e., changing the control objective to drive the generator current. However, in this case the amplitude of the load variation is unknown, hence, the generator reference is increased to $17.5$A in order to guarantee the system stability, as discussed in Section~\ref{sect:large}. Every $0.79$s $I_{OL}$ is decreased by $0.5$A, until it reaches $16$A. The result is that the current to the battery is reduced, thus compensating for the increased power demand. At time instant $t=15$s the power request is further increased by reducing the load resistor to $R_D=15\Omega$. Again, a reduced performance phase starts with  $I_{OL}=17.5$A until the nominal condition   $I_{OL}=16$A is restored within $5$s. Note that at the end of this phase the inductor current has reversed, so that actually the battery is helping the  generator. In Figure~\ref{fig:roa_Rd_IOL} the controlled state trajectory is represented in red. It is always within the ROA's. 

Figure~\ref{fig:x1s2} shows the time evolution of the inductor current, and one can note that the proposed controller tracks accurately  the current references.

Moreover, in Figure~\ref{fig:Igens2} the generated current is  reported. Note that the $5$s-capability to suppress the overload is fulfilled. 

\begin{figure}[htb]
	\centering
	\includegraphics[width=0.5\linewidth]{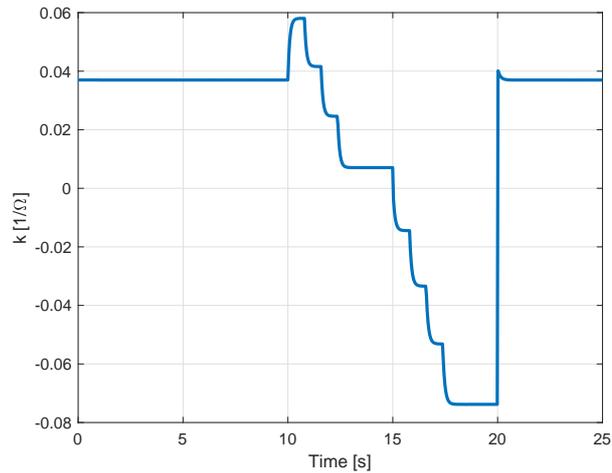}
	\caption{Adaptive parameter, Scenario 2}
	\label{fig:ks2}
\end{figure}

\begin{figure}[htb]
	\centering
	\includegraphics[width=0.5\linewidth]{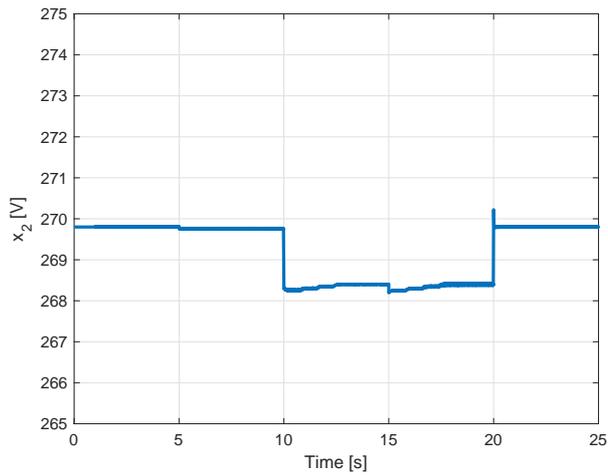}
	\caption{Capacitor $C_H$ voltage, Scenario 2}
	\label{fig:x2s2}
\end{figure}

\begin{figure}[htb]
	\centering
	\includegraphics[width=0.5\linewidth]{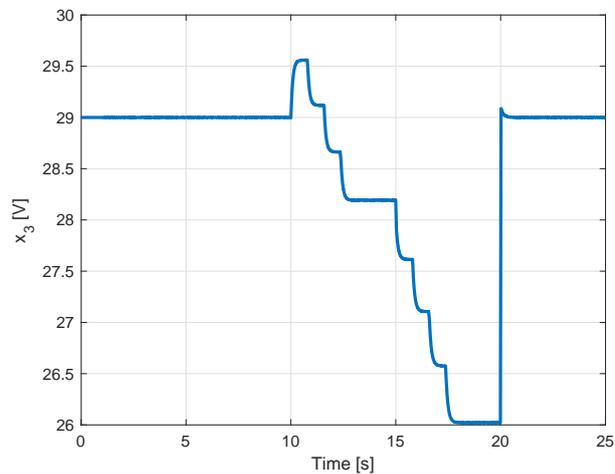}
	\caption{Capacitor $C_L$ voltage, Scenario 2}
	\label{fig:x3s2}
\end{figure}

Figure~\ref{fig:ks2} shows the time evolution of the parameter $k$. 
Finally, Figures~\ref{fig:x2s2}, \ref{fig:x3s2} show the HV and LV voltage, respectively. 

\section{Conclusions}\label{sec:conclusion}
In this paper the design of a controller for the DC/DC bidirectional converter has been discussed. The future aircraft will depend more and more on electric  devices, that must be autonomously operated. Essentially, two electrical busses are present in any aircraft, with different voltages. A DC/DC converter then is essential as a bridge between the two busses. The converter has to be controlled, and  the control actions can be selected to fulfil different objectives, e.g., to recharge the battery or to use the battery to help the generator in supplying extra-power when requested to do so. A sensible  approach is to consider two control layers, one implementing tracking and/or regulation of some variables with assured stability characteristics, the other coordinating the actions of the low-level controllers. However, it has been shown that such an approach may undergo instability, hence an integrated design has to be performed, characterising the Regions of  Attractions of the low-level controllers and ensuring that switching among different control modes can happen only when the state of the controlled system is within the ROA of the next control strategy. Thus, safe switching is ensured. Sometimes, in order to guarantee such a requirement, it can be necessary to consider a transient situation of reduced performances. All the above characteristics are discussed in details in the paper and two different scenarios are considered in a detailed simulation environment considering also switching power electronic devices, showing the effectiveness of the proposed approach. 
\section{Acknowledgement}
The authors gratefully acknowledge the contribution of Prof. Graziano Chesi for sending us the last version of the software toolbox SMRSOFT.
%



\bibliographystyle{plain}

\bibliography{automatica19}
\end{document}